# Fairness: from the ethical principle to the practice of Machine Learning development as an ongoing agreement with stakeholders

Georgina Curto
University of Notre Dame
Notre Dame, USA
gcurtore@nd.edu

Flavio Comim
Universitat Ramon Llull, IQS School of Management
Barcelona, Spain
flavio.comim@iqs.url.edu

22 March 2023

**Abstract**

This paper clarifies why bias cannot be completely mitigated in Machine Learning (ML) and proposes an end-to-end methodology to translate the ethical principle of justice and fairness into the practice of ML development as an ongoing agreement with stakeholders. The pro-ethical iterative process presented in the paper aims to challenge asymmetric power dynamics in the fairness decision making within ML design and support ML development teams to identify, mitigate and monitor bias at each step of ML systems development. The process also provides guidance on how to explain the always imperfect trade-offs in terms of bias to users.

## 1. Introduction

Discrimination and bias in AI are still unresolved topics in the current information civilization (Zuboff 2019). While AI has the potential to facilitate the achievement of all United Nations Sustainable Goals, it can also widen existing social gaps by reproducing and often aggravating societal bias (Vinuesa et al. 2020). ML systems in particular have been found to often exacerbate representational and allocational harm to vulnerable salient groups (Suresh and Guttag 2021). As a result, these groups not only receive demeaning treatment, but also less resources and opportunities. AI-amplified bias has been identified in critical services such as education, health and justice (Floridi 2020).

ML bias has other particularities that deserve special attention. Users are often not aware of it and developers cannot always explain it (Barocas & Selbst 2016). In this context, governmental efforts to regulate AI have gained traction in the past few years (White House 2016; European Commission 2021). In addition, there has been a proliferation of ethical guidelines (Algorithm Watch 2021) in what has been described as a "moral panic" (Ess 2020). These have been found to converge on specific topics (Hagendorff 2020; Zeng et al. 2019) and have been summarized as 5 ethical principles: transparency, justice and fairness, non-maleficence, responsibility and privacy (Jobin 2019).The scope of this article focuses on the principle of justice and fairness and facilitates the principle of transparency by guiding which specific steps of the fairness decision making process should be openly disclosed.

In the nascent field of AI ethics, these ethical principles have been qualified as appropriate but too abstract. AI development teams find them difficult to apply in practice. Existing AI ethics guidelines focus the effort on the "what" and fall short on clarifying the operationalization of AI ethics (Floridi 2019; Morley, Elhalal, et al. 2021; Morley, Kinsey, et al. 2021; Vakkuri et al. 2020; Vakkuri and Kemell 2019). As a result, counterproductive practices such as ethics shopping, ethics bluewashing, ethics lobbying, ethics dumping or ethics shirking are prone to flourish (Floridi 2019). An urgency has been identified to translate theoretical principles into practical inclusive processes (Harrison et al. 2020). This article has four objectives and the first one is to answer the question: how can the principle of justice and fairness be applied into the practice of ML development?

In parallel to the production of ethical guidelines, the principle of fairness has been tackled from the technical perspective by mitigating bias. A large body of work in recent years has been produced on the bias identification and debiasing of ML systems, especially on Natural Language Processing (NLP) (Bolukbasi et al. 2016; Caliskan et al. 2017; Dan Jurafskyc 2018; Guo et al. 2022; Manzini et al. 2019; Nadeem et al. 2020; Zhao et al. 2021). Intersectional biases are gradually incorporated in the analysis (Lalor et al. 2022), considering the accumulative effect of multiple biases on what Hoffman (2019) describes as the multi-oppressed groups. Approaches, such as the Dephi Project (Jiang et al. 2021) have been presented to explicitly train state-of-the-art ML models on moral judgements, weighing competing moral concerns between broad ethical norms and personal values. And gradually emphasis is changing from the quantity to the quality of "greener" data sets (Schick and Schütze 2020) and the use synthetic data that can be aligned according to specific value systems (Watson et al. 2019). However, algorithms can also be biased in the way they learn or, more appropriately, in the way they are programmed to learn (Mittelstadt et al. 2016; Tsamados et al. 2021). Additionally, AI solutions are deployed into real complex systems and it is difficult to predict the social impact of an algorithmic system before actually deploying it (Morley et al. 2020).

In spite of all the technical efforts in place to mitigate bias and the proliferation of principle-based AI ethics guidelines (Morley, Elhalal, et al. 2021), 79% of tech workers admit that they would like practical resources to assist them with ethical considerations (Miller and Coldicott 2019). It has been suggested that a more holistic method to AI bias mitigation is required, with focus not only on data and algorithms but on the procedures carried out by developers (Floridi and Taddeo 2016; Morley, Elhalal, et al. 2021). While there is an important body of work describing ML system design processes (Lehr and Ohm 2017), the quality management of such systems (Horch 1996) and the auditing practices at each stage of ML end-to-end processes development (Raji et al. 2020), to date there is not a process describing how to manage fairness decision making at each stage of ML design. This article is filling this gap.

The third objective of the paper is to incorporate the perspectives of vulnerable salient groups as regards fairness decision making within the design of ML systems (Martin et al. 2020). While participatory approaches to ML design are gaining momentum and a multi-stakeholder approach has been identified appropriate to make AI principles actionable (Stix 2021), there is a lack of clarity on how to implement inclusive AI in practice (Birhane et al. 2022). This article pursues an instrumental objective: to reach acceptable agreements among stakeholders to manage bias in a specific ML system. In addition, the paper pursues an intrinsic objective: challenge power imbalances in ML design fairness decision making. The Stakeholders Fairness Agreement (SAF) process described in the article aims to go beyond societal value alignment design (Dobbe et al. 2018; Gabriel and Ghazavi 2021). It intends to encourage reflection on societal values by providing a framework where representatives of vulnerable salient groups can express their needs and generate an impact on ML design (Sloane et al. 2020).

Finally, there is an identified need in the AI Ethics literature to analyze ML bias based on social sciences research. ML bias is often described as "statistical bias", or a mismatch between the sample to train the model and the world as it currently is (Mitchell et al. 2021). However, biases are not exclusive to the on-line world (Card and Smith 2020). When we are considering bias purely as a technical problem, we are missing part of the picture (Crawford 2017). Bias in AI should be considered a socio-technical problem (Dignum 2022). Blodgett et all (2020) analyzed 146 papers describing bias in NLP systems and explain that the proposed quantitative techniques do not engage with the relevant literature outside ML. In this article, we are offering a method to reflect on and mitigate bias in ML systems not limited to statistical bias but also tackling societal bias. For that purpose, the paper is rooted on a social sciences framework.

This paper aims to fill the identified gaps in the area of ML fairness by, first of all, providing the conceptual background outside AI to engage critically with what constitutes bias. For thar purpose, the article describes the nature of prejudice, discrimination and bias both in the online and the offline world. The paper proposes the SAF process, a hands-on, end-to-end methodology that translates the principle of justice and fairness into practice with specific actions to assist development teams at each step of the design, building, testing, deployment and monitoring of the ML lifecycle. This article does not aim to describe the ML process of ideation and development, but the process to manage bias within the ideation and development of ML systems. The SAF process foresees the identification and participation of legitimate stakeholders, including representatives of socially salient groups. As a result, it paves the way for the disclosure of the always imperfect trade-offs involved in fairness decision making and allows for a distributed responsibility of such decisions. The article discusses the challenges and limitations of the proposed process and concludes by identifying further research actions, such as assessing the suitability of the SAF process in practice through case studies, developing guidelines to evaluate compliance to it and exploring potential adaptations of the methodology to the other ethical principles and AI fields.

2. **Conceptual background: fairness as an agreement among stakeholders**

Biases have been described as systematic and predictable errors in decision making based on available heuristics (Kahneman 2011). Biases take place when there is an action, such a decision making process or the act of speech, and have their cognitive root on prejudices (Ely 1980; Greenawalt and Dworkin 1987). Allport (1954) suggests that prejudices are overgeneralized (and therefore erroneous beliefs) that lead to an attitude of favor or disfavor. Prejudices are part of the human learning process, during which we put information into categories and generalize based on previous experience. The only way to question them is by becoming aware of them through knowledge acquisition, which allows for critical thought and empathy (Cortina 2007; Morgado 2017)

**Fig. 1** Prejudices are overgeneralized and erroneous beliefs and the resulting actions can be classified according to different degrees of action (Allport 1954) which generate social stigma (Goffman 1963)

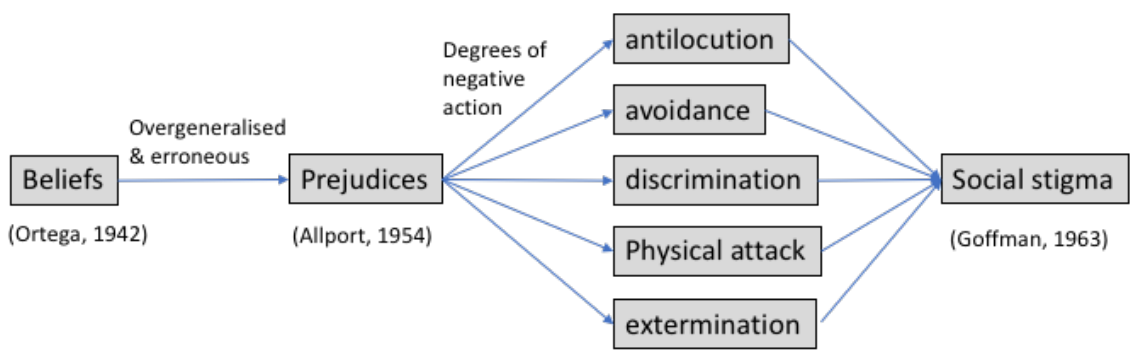

Prejudices can trigger different degrees of action defined by Allport as antilocution, avoidance, discrimination, physical attack and even extermination (Fig 1). One of the consequences of these actions is social stigma, which is associated with feelings of shame on the side of the discriminated (Goffman 1963) and beliefs of deservingness on the side of the discriminators. When prejudices are socially shared, we are talking about stereotypes. These can be transmitted through language, what we know as linguistic bias, creating a self-perpetuating cycle where prejudices are shared and maintained (Beukeboom and Burgers 2019; Maass 1999) (Fig 2).

**Fig. 2** When prejudices are shared within a specific culture we can talk about stereotypes and, when these are transmitted through linguistic bias, prejudices are reinforced in a self-perpetuating cycle

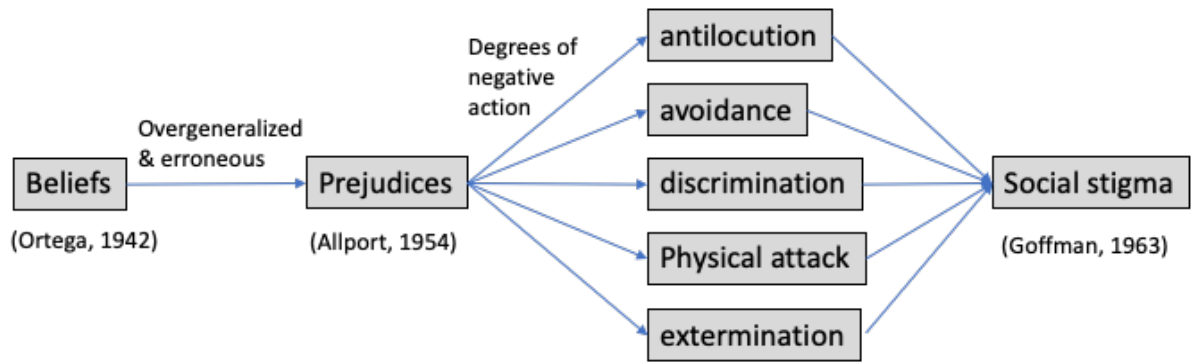

It needs to be clarified that while discrimination has become a morally-laden term (Silvers 1998), it has no build-in moral status (Eidelson 2015). In the sense of differential treatment, discrimination is a necessary concept in the legal framework when assigning rights and responsibilities such as, for example, defining a minimum age to apply for a driving license. This paper is only concerned about wrongful discrimination, which demeans the persons affected, in the sense that denies equal moral worth to individuals (Hellman 2008). Mitigating discrimination, however, should not be interpreted as impartiality or receiving equal treatment. In fact, Young describes the ideal of impartiality as keystone of a "mechanical interpretation of fairness" (1990 : 11) , which suppresses the difference that needs to be acknowledged in public policy. The mathematical approaches AI fairness currently being implemented would correspond to this mechanical interpretation. These can be grouped as: fairness through unawareness (Card and Smith 2020; Hardt et al. 2016), demographic or statistical parity (Dwork et al. 2011), individual fairness (Green and Hu 2018), randomisation (Kroll et al. 2017), equality of odds / opportunity (Hardt et al. 2016). These mathematical approximations to fairness are mutually incompatible (Card and Smith 2020; Kleinberg et al. 2016; Tsamados et al. 2021).

In fact, there is not a silver bullet to solve discrimination and bias in AI (Crawford 2017) and there is not a single universal and absolute interpretation of fairness. Young clarifies that the diversity of interpretations of fairness contains premises from the actual social context (1990). In fact, it has been argued that the widely influential conception of fairness deriving from Rawl's "distributive justice" (1971) needs to be understood in the context of liberal capitalist societies (Wolfe 1977; Iris M. Young 1981). Coeckelbergh (2022) provides some insights on how some of the main theories on fairness in social sciences could be translated to AI environments. A distributive justice approach (Rawls 1971) would require, for example, that algorithms in recruiting apps would give priority to individuals that live in worse off areas. Or according to an identitarian approach to justice (Fraser and Honneth 2003), algorithms would implement positive discrimination of vulnerable salient social groups, as described by the "algorithmic reparation" concept (Davis et al. 2021).

In this paper we argue that since ML systems cannot be subjected to a single interpretation of fairness, we propose a methodology to work towards a stakeholders' consensual view of fairness for a specific ML system. The focus should not be only on the results, but on the process to reach such consensus, where pluralism and participation of stakeholders is key (Tasioulas 2022). This approach can be found in the literature. For instance, Cortina describes fairness as "an agreement that could discover human beings through dialogue if they were really taken into account" (Cortina 2011 : 148). On his turn, Lyotard states that "there are language games in which the important thing is to listen, in which the rule deals with audition. Such a game is the game of the just" (1984 : 71) and Young explains that "rational reflection on justice begins in a hearing, in heeding a call, rather than in asserting and mastering a state of affairs" (Young 1990 : 5). The methodology described in this article seeks to provide practical recommendations to reach agreements with stakeholders in terms of fairness decision making, in order to manage the phenomenon of bias at each stage of ML design.

If ML is to benefit society as a whole, it is essential to understand the specific backgrounds of stakeholders (or agents affected by the system) (Whittlestone et al. 2019). It has to be noted that the body of work on AI bias focuses on specific demographic dimensions, mainly referring to gender and race (Bolukbasi et al. 2016; Manzini et al. 2019; Nadeem et al. 2020) or intersectional biases across multiple demographic dimensions (Lalor et al. 2022). This paper argues that these classifications are too coarse and incomplete. A more detailed analysis should be performed to identify those that are going to be adversely affected by the ML system (Stix 2021). Special attention should be paid to vulnerable salient groups, but stakeholders' representation should not be limited to these groups only. The corporation that develops and aims to use or market the AI system is responsible not only for the valueladenness of the resulting technology, but also for identifying the relevant stakeholders (Martin 2018), as it is defined in the first step of the Stakeholder Agreement for Fairness Process (SAF) (Figure 3).

In fact, one could argue that considering only the bias suffered by vulnerable salient groups could be a form of discrimination itself. Indeed, the members of dominant groups can also be victims of discrimination (even though they enjoy unfair advantages) and are therefore included in this process when they are legitimate stakeholders. However, wrongs done to persons in a dominant group are not the same as the discriminatory wrongs that combine to create serious systemic injustice ("Stanford Encyclopedia of Philosophy" 2011). Therefore, it is important to pay special attention to the groups that have been identified as being specifically vulnerable to structural discrimination when they are affected by the ML system. Which salient groups count for the purpose of determining an act of discrimination is at the heart of many political and legal debates. We are referring to vulnerable salient groups on the grounds of sex, race, color, language, religion, political or other opinion, national or social origin, association with a national minority, property, birth or other status (*European Convention on Human Rights* 2010; *International Covenant on Civil and Political Rights* 1966).

The topic of cultural specificity of the AI system in relation to the perception of fairness is also being considered. It has been documented that ML systems need to be congruent not only with the personal moral beliefs of developers, but also with the values of societies where they operate (Carman and Rosman 2021). Other proposals suggest working towards an intercultural citizenship and universal values (Jiang et al. 2021). And the ethical pluralism approach (Ess 2020; Wong 2020) acknowledges both the coexistence of universally valid values with international cultural diversity of moral codes. In practice, Chan notes that out of the top 100 universities and companies by publication index, none of them is from Africa or Latin America (2021). It is therefore essential that the stakeholders participating in the process represent the cultures where the ML system will be used, with especial attention for the inclusivity of Global South. Although stakeholders might dissent in terms of values, they all share the capacity of communicative reasoning (Habermas 1990) to reach agreements on how to manage AI fairness decision making in a specific context.

### 3. Bringing the principle of justice and fairness to the design level

The Stakeholders' Agreement on Fairness (SAF) process, shown in Fig. 3, not only aims to align ML technology with human values that receive widespread endorsement (Gabriel and Ghazavi 2021), but accompanies stakeholders on a reflective process about their own subjectivity in a specific scenario, (Terzis 2020), questioning societal values (Dobbe et al. 2018), and fosters the inclusion of a broader taxonomy of biases other than those pre-existing in the data. Therefore the SAF does not constrain the choices of stakeholders, but encourages stakeholders to make informed choices in line with the pro-ethical design concept (Floridi 2016).

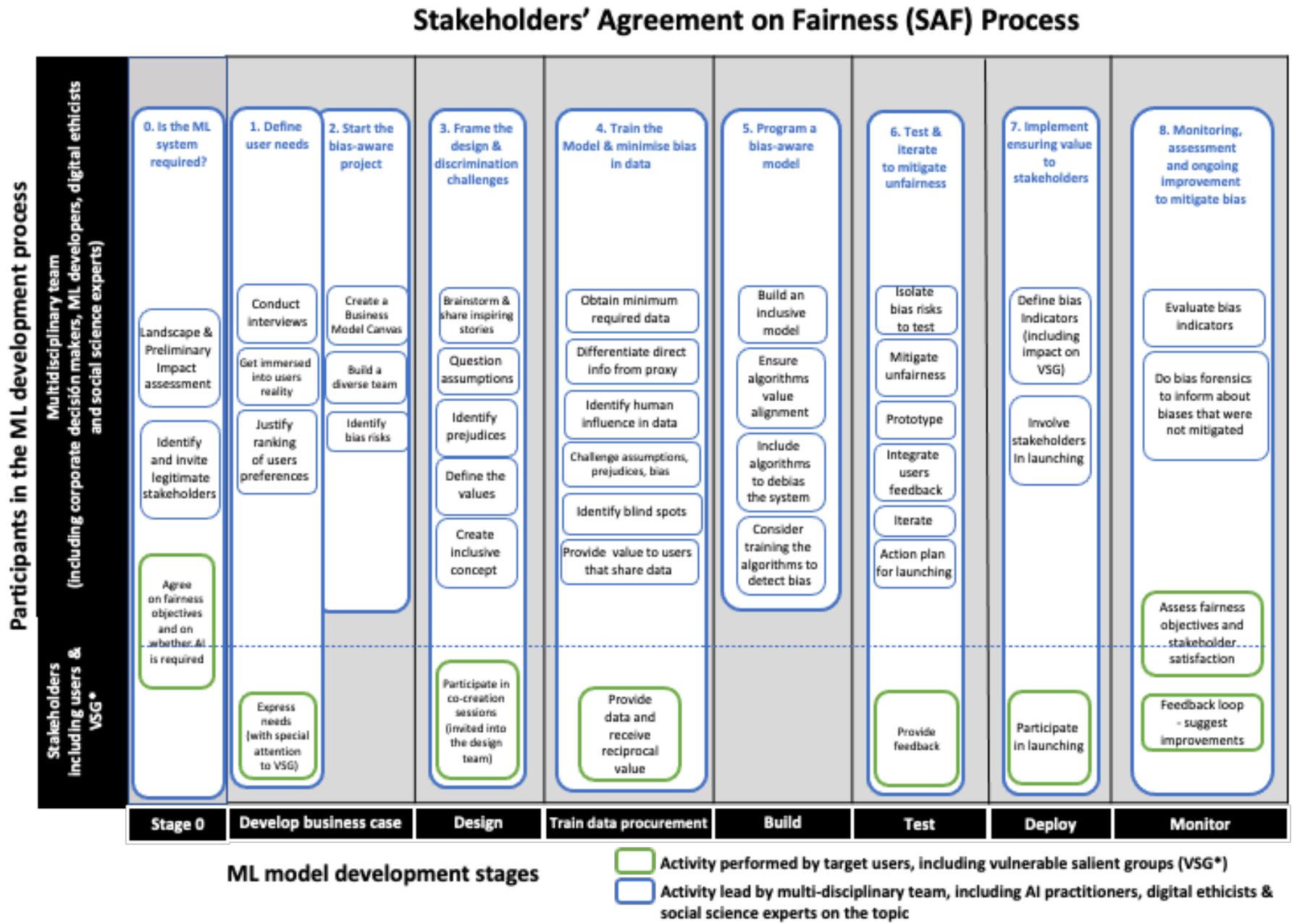

**Fig. 3** The SAF process translates the principle of justice and fairness into the practice of ML development. It is aimed at assisting AI multidisciplinary development teams to understand their own biases. Stakeholders, including vulnerable salient groups (VSG), have an active role in the process

Clarifications for each step of the SAF:

0. *Identify stakeholders, agree on fairness objectives and decide if AI is required.* As a first step, both a landscape and a preliminary impact assessment will be performed by the ML Development Multidisciplinary Team (MLMDT) before consultations with external stakeholders. The aim of the landscape assessment is to describe the contextual environment in which the ML system will be implemented (geopolitical, societal, legal) (Stix 2021). The landscape assessment will contribute to identify the stakeholders. In turn, a preliminary impact assessment will help the MLMDT identify potential risks and challenges resulting from the ML development and implementation, in that particular landscape. The preliminary impact assessment will guide the consultation process with stakeholders. Secondly, the MLMDT needs to identify the legitimate stakeholders, i.e., all the agents affected by the ML system (both internal and external to the company or institution developing it), including vulnerable salient groups. Within the stakeholders' group is important to balance the need for subject-matter technological expertise with the diversity of perspectives and to manage power imbalances (Hollis and Whittlestone 2021). Once the stakeholders' representatives have been identified, the aim of the stage 0 of the SAF process is to question whether the project goals contribute to the human objectives shared by the stakeholders. An explicit agreement on what are the objectives of the ML system in terms of fairness should be reached among stakeholders. As a result of stage 0, stakeholders must feel empowered to conclude that the ML system is not required at all, in which case it should not be developed (Pasquale 2019; Russell 2019). Corporative objectives such as efficiency, performance, accuracy, novelty and state-of-the-art are to be questioned and societal benefits according to the requirement of "diversity, non-discrimination and fairness" (HLEGAI 2019 : 14) should be agreed and assessed at the end of the ongoing process when the outputs of the project are obtained (Hollis and Whittlestone 2021).

1. *Define users' needs (including those of vulnerable salient groups (VSG)).* Once stakeholders are identified, the MLMDT will focus on documenting the users' needs that the ML system aims to address, which are key to human-centered design (IDEO.org 2015). The deeper the MLMDT gets into the users' reality (including VSG), the more it will be able to understand users' beliefs and values and therefore question social assumptions and prejudices. There are trade-offs in all development processes and the MLMDT needs to justify the ranking of users' preferences in order to provide explainability.

2. *Start the bias-aware project.* Building a diverse team is an integral part of the project in order to achieve ethical pluralism (Ess 2020). And the practical operationalization of AI ethics is not about external impositions, but more about procedural regularity (or phronesis). Therefore, the process aims to accompany the MLMDT to continuously learn from own subjectivity and biases, adapting the process across contexts, and reach agreements with stakeholders (Kroll et al. 2017). The MLMDT should identify the bias risks, which should be taken into account in the business model canvas.

3. *Frame the design & discrimination challenges.* The needs of the VSG should be taken into account in the brainstorming sessions, paying special attention to personal experiences. Having detailed information on users (including real-life situations of exclusion) provides knowledge on other cultures and contexts that help identify values, assumptions and counteract prejudices. Stakeholders should be invited into the design team in co-creation sessions (IDEO.org 2015), in line with the Trustworthy AI requirement of "human agency and oversight" (HLEGAI 2019 : 15) and the capability approach to agency (Nussbaum 2012; Sen 2001).

4. *Train the model and minimize bias in data.* Enormous amounts of data tend to include low quality information and higher level of biases. Therefore, the focus should not only be on obtaining the maximum quantity of data, but on ensuring the maximum quality of this data, which will often mean working with smaller datasets (Schick and Schütze 2020). Data needs to be analyzed to challenge assumptions, prejudices and the resulting bias by differentiating direct information from proxy, identifying human influence as well as blind spots (Sampson, O., & Chapman 2021). Existing approaches to identify and measure bias in data can be explored (Bolukbasi et al. 2016; Dan Jurafskyc 2018; Kiritchenko et al. 2014; Manzini et al. 2019; Nadeem et al. 2020; Zhao et al. 2021). In addition, the MLMDT needs to bear in mind that value should be provided to users that share data in order to comply with fairness criteria.

5. *Program a bias-aware model.* Data from the "real world" cannot be assumed to have the values agreed within the SAF process. Therefore, algorithms to debias the system are to be foreseen (Bolukbasi et al. 2016; Manzini et al. 2019; Zhao et al. 2021). The development team can also consider using methods to train algorithms to detect bias (Jiang et al. 2021; Sap et al. 2020).

6. *Test & iterate to mitigate unfairness*. Identified bias risks are to be tested in isolation to ensure unfairness is mitigated. Users' feedback can be integrated into several iterations of the prototype testing, in line with the concept of non-bias engineering of negotiated ethics (Morley, Elhalal, et al. 2021). In addition, emergent bias in the system should be identified by studying the feedback mechanisms between the algorithms and the environment they act upon (Dobbe et al. 2018).

7. *Implement ensuring value to stakeholders*. Indicators are to be defined in order to measure the impact of the ML model on stakeholders and monitor the achievement of agreed objectives on fairness and bias mitigation. This information needs to be publicly available in the launching of the ML model and thereafter, in line with the transparency principle. Target users and other stakeholders, including VSG, are invited to participate in the launch validation and are solicited feedback after deployment, in line with the "stakeholder participation" approach recommended in the EU Ethics Guidelines for Trustworthy AI (HLEGAI 2019: 19).

8. *Monitoring, assessment and ongoing improvement to mitigate bias*. AI ethics focuses on procedural regularity (Morley, Elhalal, et al. 2021) and the SAF process is not to be applied as a "one-off" test, but to be re-applied on ongoing basis when ML systems are revised and re-tuned. The agreed objectives in terms of fairness should be revisited and enriched in line with the critical maturity of society. Consultation to stakeholders post-implementation is foreseen with a feedback loop to ensure continuous process improvement (Stix 2021). The SAF is an ongoing process that will evolve and keep track of biases in a context of its time. Since not all biases will be eliminated, the MLMDT will need to do bias forensics (Crawford 2017), to be able to inform on biases in an open and transparent way. For the SAF process to be effectively integrated in the ML development practices of a company or institution, it is recommended to implement it first on a pilot project where there are clearly identified stakeholders. This will allow to build the in-house expertise to be able to address more ambitious projects (Whittlestone and Clark 2021).

## 4. Facilitating trade-offs disclosure

The SAF process creates the grounds to openly disclose the agreements reached on the principle of justice and fairness. Since decisions are reflected upon and agreed among stakeholders, they are easier to communicate. The SAF process, therefore, facilitates the Trustworthy AI requirement of "transparency, including traceability, explainability and communication" (HLEGAI 2019:14).

Transparency has been described as a second-order principle because it can be directly addressed from a programming perspective, tackling the black-box effect (Carman and Rosman 2021; Floridi et al. 2018). It allows organizations to communicate the always imperfect trade-offs (Whittlestone et al. 2019). In fact, it is argued that when a system is explainable and interpretable it is inherently fairer, since it allows stakeholders take informed decisions on whether to use the ML system (Binns et al. 2018; Cath 2018; Lipton 2016). AI systems need to be designed to be transparent (Ananny and Crawford 2018) and Figure 4 defines what is the minimum information from the SAF process that should be explicitly communicated in order to disclose the fairness decisions taken throughout the development of the ML system.

**Fig. 4** Description of the minimum information throughout the SAF process that needs to be disclosed so that users can take ethically informed decisions on whether or not to use the system, in line with the transparency requirement (HLEGAI 2019 : 18)

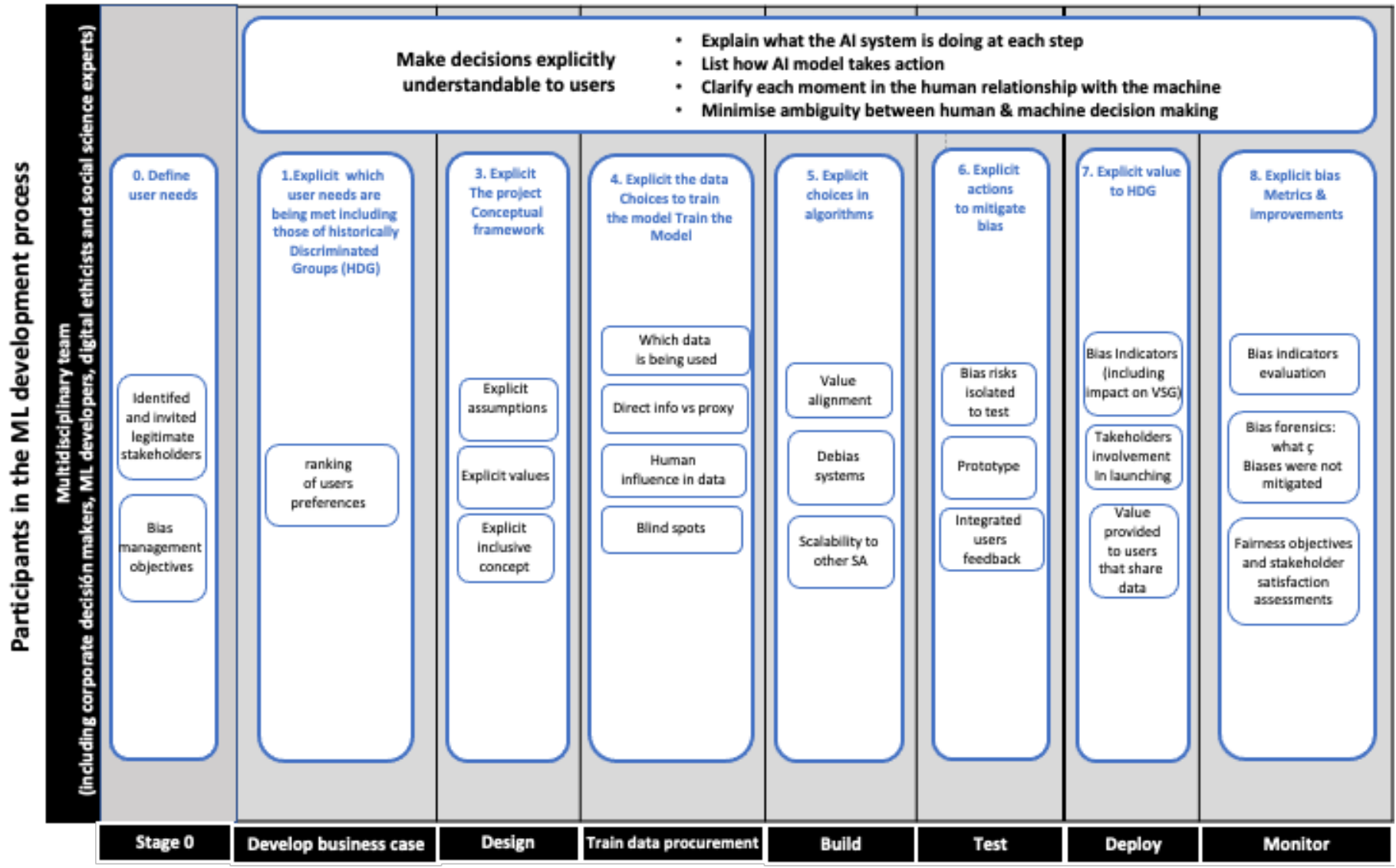

## 5. Challenges and limitations

The SAF process is grounded on the state on the topic of fairness both in AI and social sciences. However, it has not yet been empirically validated. Testing it in a real context scenario will be required to prove its applicability, ensure the benefits on bias mitigation and improve the drawbacks. Secondly, all approaches aiming to tackle complex ethical principles have some limitations and this is not an exception. Table 1 identifies the main challenges and risks as well as the proposed mitigating actions.

Table 1. Challenges and risks of the SAF process and proposed mitigating actions

| Identified challenge / risk | Description | Mitigating action |
|---|---|---|
| Lack of representation of all agents affected | Martin and Phillips (2022) explain that a collection of social, legal and economic forces drive firms to prioritize and reinvest in current stakeholders. | We have explicitly stated, at stage 0 of the SAF process, that all legitimate stakeholders should be represented, i.e., all agents affected by the AI system, both internal and external to the company or institution that is developing it. |
| Power imbalances among participating stakeholders | Opinions expressed by vulnerable salient groups might be undervalued by AI powerful stakeholders (Birhane et al. 2022). This can aggravate rather than improve the existing power dynamics. | The MLMDT must discuss how to deal with power imbalances before inviting stakeholders to the project. If necessary, the team will look for expert advice to ensure that participants feel free and empowered to express their opinions (Hollis and Whittlestone 2021). Tensions among stakeholders are bound to arise, since fairness is not achieved in effortless ways (Sloane et al. 2020). The MLMDT aims to accompany stakeholders throughout the process to reach agreements that are acceptable by all stakeholders. |
| Participation washing | Participatory processes can be a mechanism for big corporations to mask or even facilitate an illegitimate exercise of power, focusing only on corporate benefits. In this case, the disempowerment of VSG continues and corporations emerge as the legitimate arbiters (Sloane et al. 2020). | Objectives in terms of bias management and participation will be agreed among legitimate stakeholders, openly published and assessed periodically. Assessment on stakeholders' participation satisfaction is also required in the SAF process. Participation will only be considered effective if the agreed objectives on bias management are met and legitimate stakeholders express a positive outcome in the participation assessment. |
| Lack of clarity on what meaningful participation entails | If specific steps and objectives of the process are not clearly defined within the design, development and deployment steps, the good intentions of participation can be diluted and not fructiferous. | The SAF proposes a step-by-step process identifying the participation of stakeholders and the agreement on objectives that are evaluated periodically. |
| A cover up for unethical or illegal activities | If there is not an evaluation of the SAF process, there is a lack of assurance that it complies with AI Trustworthy requirements (HLEGAI 2019) and the legal framework. The SAF process can question current legislation and become a trigger to modify it, but it should not justify going against the law. | Guidelines to perform an impact assessment of the SAF process implementation should be proposed. These would provide organizations a way to anticipate the potential effect of SAF in terms of outputs. Guidelines to evaluate compliance with the SAF process should also be developed building on the vast literature on methodologies of AI assurance, especially to audit AI systems (Ahamat et al. 2021; Asplund et al. 2020; Metaxa et al. 2021) (Raji et al. 2020). In turn, SAF process is expected to facilitate existing processes of AI auditing, certification and |

|  |  | accreditation because it provides explainability of the fairness decision making. |
|---|---|---|
| Conflation of the process | The participation of legitimate stakeholders does not mean that there are no discriminatory outcomes. | At a minimum, the outcome of the SAF process should comply with the criteria of demographic parity and negative dominance (Wachter et al. 2020). For example, black applicants should not make for the majority of rejected applicants in job recruitment. In addition, the always imperfect trade-offs in terms of fairness decision making are to be disclosed. |
| Difficulty to measure the benefits of participation | As a result, the positive benefits of participation can be questioned in terms of profitability. | Participation should not be considered the means to a goal, but also a goal in itself (Birhane et al. 2022), where stakeholders feel empowered to evaluate the impact of the technical solution and even to conclude that the solution is not required at all. A participation satisfaction assessment is foreseen in the SAF process to measure the benefits of the process. |
| Burden on all stakeholders to gather information, reflect and decide | Reflecting and taking decisions on bias can become an enormous task. | Reflection and decision making in the SAF process is limited to a specific scenario and stakeholders will be assisted by the project MLMDT to focus on the specificity of such context. |
| Lack of incentives for engaging in the process | The SAF process might be seen as an overhead by the companies and institutions that develop and implement ML systems. | There is an increasing pressure from employees and users to implement fairer and more transparent ML systems. 63% of technological workers declare that they need more time and resources to think about the impact of their work (Miller and Coldicott 2019). Therefore, the principle of AI justice and fairness is increasingly a business need. |
| Lack of representation of stakeholders from the Global South | Two major machine learning venues, NeurIPS 2020 and ICML 2020, found that among the top 10 countries in terms of publication index, none are located in Latin America, Africa or the Southeast Asia (Chuvpilo 2020).<br><br>There is a risk, therefore, that stakeholders from the Global South, where ML systems are also applied, are not invited to participate in the process. | The SAF process precisely focuses on involving the legitimate stakeholders where the ML system is to be implemented. |
| Design justice as an oxymoron | Given the logic of performance focus and profit orientation in which ML systems are developed and implemented, obtaining ML products that are genuinely fair and equitable can be considered an impossible task not worth pursuing (Sloane et al. 2020). | The SAF process does not aim to provide universally fair outputs, but to reach acceptable agreements among legitimate stakeholders who, feeling empowered to do so, define and assess objectives in terms of fairness for a particular scope and ML developing process. |

## 6. Conclusions

In recent years numerous studies have acknowledged that ML systems can have harmful consequences in terms of human rights and discrimination. A growing number of voices describe a need to translate the ethical principle of justice and fairness into practice and call for the participation of social science researchers to clarify and contextualize the concept of bias. To fill the existing gaps, this paper provides, first of all, a descriptive framework for the concepts of prejudice, discrimination and bias. Since prejudices are originated in the way human beings interpret reality, bias cannot be mitigated completely, rather it should be managed not only in the data and algorithms but also in the practices of ML development. With that aim, the SAF process constitutes an end-to-end inclusive framework that encourages an ongoing reflective approach to bias management and mitigation, by suggesting specific actions to be taken by a multi-disciplinary team and active involvement of stakeholders, including VSG. As a result, the SAF process facilitates the disclosure of the trade-offs when managing bias, giving users the necessary information to take ethically-informed decisions. In addition, the transparency provided by the SAF process facilitates the external assessment, because it provides explainability on the decisions taken (Dearden and Rizvi 2008). Therefore, the SAF process can constitute a useful tool for NGOs, community organizations and government officials to monitor ML and encourage its alignment with broad civic goals rather than narrow commercial interests (Whittlestone and Clark 2021).

Societies where ML systems operate are becoming better informed and more critically aware about the challenges put forward by topic of AI ethics. Stakeholders expect ML systems to benefit the communities where they operate and AI ethics is growingly becoming a business need. However, socially beneficial ML systems cannot be achieved as one-shot activity nor with technical solutions exclusively, but rather be the result of procedural regularity and inclusive participation. Further work should be performed to verify the suitability of the SAF process in practice, through case studies. A multi-disciplinary ethics advisory board should evaluate the appropriateness and comprehensiveness of the SAF process (Morley, Elhalal, et al. 2021) and guidelines should be developed to evaluate compliance with it. Finally, further research should explore potential adaptations of the SAF process to other Trustworthy AI requirements and broadening the scope to other AI fields such as robotics.